\documentclass{aa}  

\usepackage{graphicx}
\usepackage{txfonts}
\usepackage{orcidlink}
\usepackage{newtxtext,newtxmath}
\usepackage{booktabs}
\usepackage{upgreek}
\usepackage{threeparttable}
\usepackage[T1]{fontenc}
\usepackage{comment}
\DeclareRobustCommand{\VAN}[3]{#2}
\let\VANthebibliography\thebibliography
\def\thebibliography{\DeclareRobustCommand{\VAN}[3]{##3}\VANthebibliography}
\usepackage{graphicx}	
\usepackage{amsmath}	
\usepackage{tablefootnote}
\usepackage{url}
\bibpunct{(}{)}{;}{a}{}{,} 

\usepackage{hyperref}
\hypersetup{
    colorlinks=true,
    linkcolor=blue,
    citecolor=blue,
    filecolor=blue,
    urlcolor=blue     
}

\begin{document}

\title{Deep polarimetry study reveals double ring ORC-like structures}

   \author{Sam Taziaux \inst{\ref{rub},\ref{rapp}, \ref{csiro_perth}}\orcidlink{0009-0001-6908-2433}
          \and
           Dominik J. Bomans \inst{\ref{rub}, \ref{rapp}}\orcidlink{0000-0001-5126-5365}
           \and
           Christopher J. Riseley\inst{ \ref{rub}}\orcidlink{0000-0002-3369-1085}
         \and
           Alec J. M. Thomson \inst{\ref{csiro_perth}}\orcidlink{0000-0001-9472-041X}
         \and
            Ray P. Norris \inst{\ref{csiro_sydn},\ref{wsu}}\orcidlink{0000-0002-4597-1906}
        \and
          Aritra Basu \inst{\ref{TLS}, \ref{mpi_bonn}}\orcidlink{0000-0003-2030-3394}
         \and 
           George H. Heald \inst{\ref{csiro_perth},\ref{ska}}\orcidlink{0000-0002-2155-6054}
        \and 
         Timothy J. Galvin \inst{\ref{csiro_perth}}\orcidlink{0000-0002-2801-766X}
        \and 
        Björn Adebahr \inst{\ref{rub}}\orcidlink{0000-0002-5447-6878}
        \and
           Miroslav D. Filipovi\'c \inst{\ref{wsu}}\orcidlink{0000-0002-4990-928}
        \and Nikhel Gupta \inst{\ref{csiro_perth}}\orcidlink{0000-0001-7652-9451}
        \and Stas Shabala \inst{\ref{tasmania}}\orcidlink{0000-0001-5064-0493}
        \and Tayyaba Zafar 
        \inst{\ref{macquarie}}\orcidlink{0000-0003-3935-7018}
          }

   \institute{Ruhr University Bochum, Faculty of Physics and Astronomy, Astronomical Institute (AIRUB), Universitätsstraße 150, 44801 Bochum, Germany
              \email{sam.taziaux@rub.de} \label{rub}
        \and 
             Ruhr Astroparticle and Plasma Physics Center (RAPP Center) \label{rapp}
         \and 
            CSIRO Space and Astronomy, PO Box 1130, Bentley WA 6102, Australia \label{csiro_perth}
        \and 
             Australia Telescope National Facility, CSIRO Space and Astronomy, PO Box 76, Epping, NSW 1710, Australia \label{csiro_sydn}
        \and 
             Western Sydney University, Locked Bag 1797, Penrith South DC, NSW 2751, Australia  \label{wsu}
        \and
             Th\"uringer Landessternwarte, Sternwarte 5, 07778 Tautenburg, Germany \label{TLS}
         \and
             Max-Planck-Institut für Radioastronomie, Auf dem H\"ugel 69, 53121 Bonn, Germany \label{mpi_bonn}
             \and 
              SKA Observatory, SKA-Low Science Operations Centre, 26 Dick Perry Avenue, Kensington WA 6151, Australia \label{ska}
              \and
             School of Natural Sciences, University of Tasmania, Private Bag 37, Hobart TAS 7001, Australia. \label{tasmania}
             \and 
             School of Mathematical and Physical Sciences, Macquarie University, NSW 2109, Australia \label{macquarie}
            }

\date{Accepted XXX. Received YYY; in original form ZZZ}



\abstract
{New observations with the current generation of advanced radio interferometers, such as ASKAP and MeerKAT, have allowed observers to discover new classes of extended radio sources of unknown origins, including the 'Odd Radio Circles' (ORCs). These phenomena are detected exclusively in the radio continuum, with no clear counterparts of the structures at other wavelengths, making their physical nature and origin a subject of ongoing investigation.}
{To better understand these phenomena and their origin, we study their radio continuum emission, spectral characteristics, and magnetic field properties.}
{This study presents a radio spectropolarimetry analysis of a newly discovered ORC (ORC\,J0356--4216) displaying a rare double-ring morphology. We use data from the MeerKAT L-band and from the ASKAP Evolutionary Map of the Universe (EMU) at 943\,MHz.}
{ORC\,J0356--4216 shows a symmetric double ring structure with a diameter of approximately $2\arcmin$, corresponding to a physical size of 667.6\,kpc, based on the redshift ($0.494\pm0.068$) of its apparent host galaxy WISEA\,J035609.67--421603.5. The radio spectra of both rings are steep, with spectral indices of $-1.18\pm0.03$ and $-1.12\pm0.05$, and show no significant substructures. Equipartition magnetic field strengths with $K_0=1$ are estimated to be $1.82\,\upmu$G and $1.65\,\upmu$G for the respective rings. The degree of polarisation across the object ranges between 20–30\,\%, further supporting a non-thermal origin.} 
{The morphology and polarisation are broadly consistent with large-scale shocks driven by powerful starburst outflows. Nevertheless, the high degree of symmetry, the coherent double-ring structure, and the absence of internal substructures are features commonly associated with relic AGN lobes, making this scenario particularly compatible with the observed characteristics.}


   \maketitle
%




\section{Introduction}
Odd Radio Circles (ORCs) represent a newly discovered morphological class of extragalactic objects characterised by their distinctive ring-like morphology in the radio wavelengths of the electromagnetic spectrum \citep[e.g.][]{Norris_2021, Koribalski_2021,Norris_2025,Gupta_2022,Gupta_2025}. ORCs typically have diameters of the order of about one arcminute, corresponding to physical sizes of several hundred kiloparsecs.  In several cases, there is evidence of a fainter second ring \citep[e.g.][]{Norris_2022, Riseley_2024}.

These structures were first identified in the first pilot survey~\citep[EMU-PS1,][]{Norris_2021b} for the Evolutionary Map of the Universe \citep[EMU,][]{Hopkins_2025} survey using the Australian SKA Pathfinder \citep[ASKAP,][]{Hotan_2021}. 
The first ORCs \citep[][]{Norris_2021a} included ORC\,1 (ORC\,J2103--6200) which is centred on an elliptical galaxy, ORCs\,2/3 (ORC\,J2058--5736 which are associated with a double-lobed AGN, but are now not considered to be a {\em bona fide} ORC, and ORC\,4 (ORC\,J1555+2726) \citep[][]{Norris_2021,Riseley_2024, Coil_2024} which was found in GMRT data. ORC\,5 (ORC\,J0102–-2450) was later found in ASKAP data \citep[][]{Koribalski_2021}. A much smaller ORC (ORC\,J0219--0505) of around 33\,kpc has been discovered in the MIGHTEE survey with MeerKAT \citep[][]{Norris_2025}. There have been some more ORC detections, such as ORC\,J2223--4834 \citep[][]{Gupta_2022}, ORC\,J0210--5710, ORC\,J0402--5321, ORC\,J0452--6231, ORC\,J1313--4709 and ORC\,J2304--7129 \citep[][]{Gupta_2025}.
All these ORCs have host galaxies sharing similar properties, being massive ($>10^{11}\,{\rm M}_\odot$) and old \citep[$>1\,$Gyr;][]{Rupke_2024,Coil_2024}.

While the original definition of ORCs was relatively broad,  a tighter definition has recently been adopted to avoid confusion with other phenomena. \citet{Norris_2025} defines an ORC as an edge-brightened circle of radio emission, without any known corresponding emission at other wavelengths, surrounding a distant galaxy. 
In addition to the ORCs, other circular objects have been reported in various environments, which may share some characteristics with ORCs \citep[e.g.][]{Bordiu_2024, Lochner_2023, Koribalski_2024a, Koribalski_2024b, Dolag_2023, Gupta_2022}. 
They exhibit a wide range of morphologies, sizes, environments, and physical properties. For instance, the `Physalis' system (ASKAP J1914–-5433) features large radio shells surrounding an interacting pair of galaxies. This indicates that such radio features can form in dynamically active environments \citep[][]{Koribalski_2024b}.
Galactic objects have also been found that resemble ORCs, but are orders of magnitude smaller (few kpc), such as the recently discovered structure around the Galactic centre named K\'yklos. This shows flatter spectral indices ($\alpha = -0.12 \pm 0.56$), which could be consistent with thermal Bremsstrahlung \citep[][]{Bordiu_2024}. K\'yklos may be an evolved supernova remnant (SNR) in an extremely low-density medium, with the emission possibly having a significant thermal component \citep[][]{Bordiu_2024, Sarbadhicary_2023}. Another recently-found example is the `Anglerfish' (Filipovi\'c et al. in prep.).  

A key characteristic of ORCs is the absence of corresponding diffuse emission at other wavelengths, such as optical, infrared, or X-ray wavelengths \citep[e.g.][]{Norris_2022, Koribalski_2024b}.  However, a notable characteristic of the known ORCs is that they have massive elliptical galaxies located near their centre \citep[e.g.][]{Koribalski_2021, Norris_2021,Norris_2021a, Rupke_2024, Fujita_2024,Lin_2024, Shabala_2024,Yamasaki_2024, Coil_2024}. 
All these host galaxies share very similar properties, suggesting a potential physical connection between the host galaxy and the surrounding radio structure. 

The radio emission from ORCs is generally weak and diffuse, with total flux densities ranging from 1 to 5\,mJy at 1\,GHz. Spectral index analysis of this emission suggests a non-thermal origin, especially with steep negative values ($\alpha < -0.4$), where the flux density ($S\propto \nu^\alpha$). This points to synchrotron radiation from relativistic electrons as the dominant emission mechanism \citep[e.g.][]{Norris_2022}.  
Polarisation measurements of ORCs also indicate the presence of ordered magnetic fields within the emitting regions, consistent with a shell of emission \citep{Norris_2022}.

The physical nature and formation process of ORCs remain a subject of intense research and discussion. Several scenarios have been proposed to explain their origins, including 
(1) spherical shockwaves triggered by energetic, transient events in galaxies, such as the mergers of supermassive black holes \citep[e.g.][]{Yamasaki_2024, Norris_2021}, or galaxy mergers \citep[e.g.][]{Dolag_2023, Yamasaki_2024, Koribalski_2024b}, 
or by shocks from explosive episodes of star formation, or a starburst-driven shockwave \citep[e.g.][]{Norris_2021, Coil_2024, Yamasaki_2024}. These shocks could accelerate relativistic particles in the surrounding intergalactic or circumgalactic medium, leading to the observed synchrotron emission; 
(2) the hypothesis of remnants of galactic outflows (Outflowing Galactic Radio Emitting Structures, OGREs), where energetic matter ejections from the centres of galaxies produce shockwaves that accelerate cosmic rays \citep[][]{Fujita_2024}; 
(3) several systems exhibiting diffuse radio emission surrounding distant galaxies, potentially representing early stages of ORC evolution or alternative evolutionary paths, referred to as GLAREs (Galaxies with Large-scale Ambient Radio Emission) \citep[e.g.][]{Gupta_2022,Gupta_2025});
(4) the scenario in which ORCs represent reactivated radio lobes from previous phases of AGN activity, as the result of shocks from mergers of galaxy clusters or groups \citep{Shabala_2024};
(5) the possibility that synchrotron emission from relativistic electrons accelerated in cosmological virial shocks within galaxy halos could explain the observed properties of ORCs \citep[][]{Yamasaki_2024}; and 
(6) the evolution of AGN jet-inflated bubbles by injecting mass \citep{Lin_2024}.

Here we report the discovery of a new ORC (ORC\,J0356--4216),  and perform a polarimetry study to explore the mechanisms responsible for the ring-like structure and draw conclusions on their origin and nature.
We describe our data in Sect.~\ref{data}. The properties of the ORC itself, as its radio continuum properties and polarisation properties are described in Sect.~\ref{results}. In Sect.~\ref{discussion}, we place our observations in context with other ORCs and make deductions about their natures. In Sect.~\ref{conclusion}, we present a summary of our work and our conclusions. 
\begin{figure*}
    \centering
    \includegraphics[width=\linewidth]{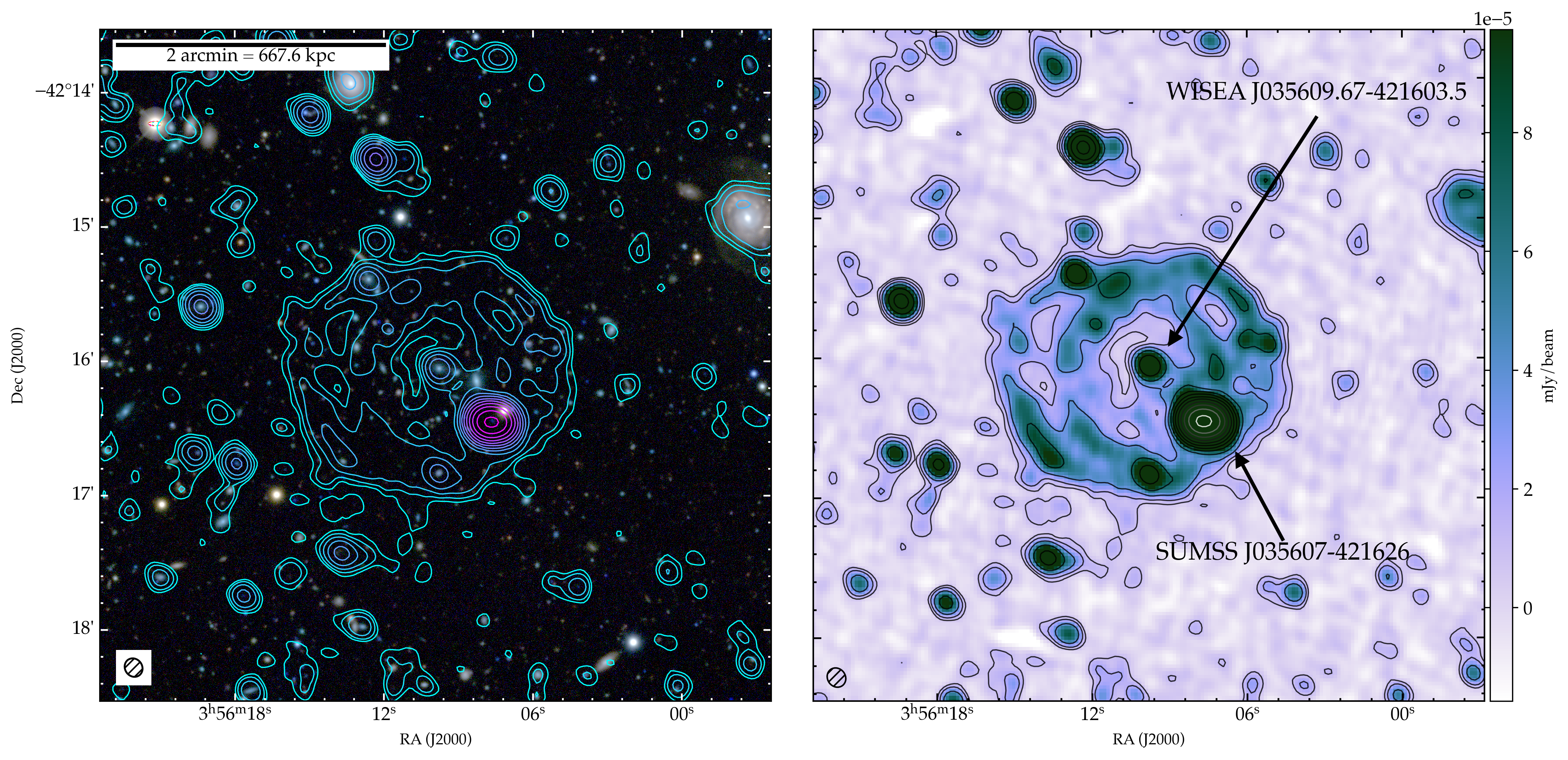}
    \caption{\textit{Left:} Color-composite image from the DESI Legacy Imaging Surveys of ORC\,J0356–4216 with overlaid MeerKAT L-band radio emission contours starting at $3\,\sigma$ and increasing by a factor of 2 at a central frequency of 1.28\,GHz with a noise level $\sigma=3\,\upmu$Jy/beam. The scale in the top left corner is calculated using the redshift of the host galaxy WISEA\,J035609.67--421603.5. 
    \textit{Right:} MeerKAT L-band radio emission with overlaid contours starting at $3\,\sigma$ and increasing by a factor of 2 at a central frequency of 1.28\,GHz with a noise level $\sigma=3\,\upmu$Jy/beam. The arrows are pointing to the host galaxy WISEA\,J035609.67--421603.5 and the bright radio source SUMSS\,J035607--421626.
    The beam of $8.57\arcsec \times 7.65\arcsec$ is shown in the bottom left corner.}
    \label{desi}
\end{figure*}
\section{Observations and data reduction}
\label{data}
\subsection{MeerKAT}
\subsubsection{Observation}
ORC\,J0356--4216 was in observations performed on 12-October-2023 and 15-October-2023 as part of a radio astronomy projects using the MeerKAT telescope (Project ID: SCI-20230907-ST-01, PI: S. Taziaux). The observations were conducted in the L-band (1.28\,GHz) and covered a total integration time of 6\,hrs. Although ORC\,J0356--4216 was not the primary target of these observations, it lies at an angular distance of $6.68\arcmin$ from the phase centre of the observed field for each observation, which is still (1) within the primary beam FWHM of MeerKAT L-band and (2) within the radius where MeerKAT's off-axis polarisation is negligible \citep[][]{Sekhar_2022}.

\subsubsection{Data reduction}
\label{datameerkat}
 The MeerKAT observations at L-band, covering the frequency range 856 to 1712\,MHz, were recorded using 32768 frequency channels. These data were
averaged to 4096 209.984\,kHz-wide channels for calibration at a central 
frequency of 1284\,MHz. On both days of observations, 
J$0408\textrm{--}6545$ was used to set the
absolute flux density scale, solve the bandpass, and because it is weakly polarisation,
it was also used to solve the on-axis instrumental leakage. The complex gains were solved
using an unresolved bright source, J$0440\textrm{--}4333$, and the absolute polarisation
was calibrated using 3C\,138. Calibration were performed following standard
procedure using the \texttt{CASA package} \citep[][]{casa}, wherein, the absolute flux scale for J$0408\textrm{--}6545$ was set using the task \texttt{setjy} following the strategy provided by the SARAO,\footnote{\url{https://skaafrica.atlassian.net/wiki/spaces/ESDKB/pages/1481408634/Flux+and+bandpass+calibration}} and the polarisation
model for 3C\,138 was adopted from \cite{PerleyButler2013}.

Iterative imaging and self-calibration using \textsc{ws-clean} \citep{wsclean} and \texttt{CASA}, (phase-only, frequency-independent, with a solution interval of 2.5\,min down to 10\,s) until image quality reached convergence with a Briggs weighting with $\texttt{robust} = -1, -0.5, -0.3, 0, 0.3$ to slowly reconstruct the diffuse emission. 
Self-calibration was performed independently for the two days of observations, and by splitting the 856-MHz bandwidth into 16 sub-bands of 53.5\,MHz each, to avoid bandwidth smearing larger than the synthesized beam, during imaging of the wide field of view.
Multifrequency and multiscaling CLEANing \citep{hoegbom_1974} utilising interactive masks using \texttt{flint\_masking}~(Galvin et al. in prep.)\footnote{\url{https://github.com/flint-crew/flint}} around visible sources to minimise artefacts and flux scattering and applying detection thresholds to retain genuine emission only.
After satisfactory solutions were obtained using the target field, 
final imaging in all Stokes parameters were done by combining the two datasets in 
\textsc{ws-clean}.
The total intensity continuum image was generated using Briggs robust weighting with $\texttt{robust} = 0$, resulting in resolutions of $8.57\arcsec \times 7.65\arcsec$, and rms of $3\,\upmu$Jy/beam limited by the confusion noise. 
The primary beam have been corrected using \texttt{-apply-primary-beam} task in \textsc{ws-clean}.

For polarisation imaging, we individually imaged the Stokes-$Q$ and $U$ parameters with a channel width of 13.5\,MHz to mitigate bandwidth depolarisation effects at a Briggs robust weighting of $0.5$ to trace the diffuse polarised emission. We used a higher positive robust weighting parameter for the Stokes-$Q$ and Stokes-$U$ imaging than for Stokes-$I$ in order to enhance the sensitivity to diffuse polarised emission. The polarised intensity ($PI$) was derived as the maximum value along the Faraday axis of the polarised intensity cube, calculated from $PI = \sqrt{Q^2 + U^2}$ \citep[][]{Wardle_1974}. 
We use some additional parameters during polarisation imaging with \textsc{WS-clean}, such as \texttt{-squared-channel-joining}, \texttt{-join-polarizations} and \texttt{-squared-channel-joining} to optimise the quality of the result.
The resulting 42 images have been primary beam corrected and convolved to lowest common resolution of $22\arcsec$ to perform rotation measure (RM) synthesis \citep[][]{Burn_1966, Brentjens_2005, Heald_2009_II} using the \textsc{RMtools}\footnote{\url{https://github.com/CIRADA-Tools/RM-Tools}} \citep[][]{rmtools} suite task \texttt{rmsynth3d}, sampling between $-700$ and $700\,\,\mathrm{rad}\,\mathrm{m}^{-2}$ with a stepsize of $4\,\,\mathrm{rad}\,\mathrm{m}^{-2}$. We do not perform \texttt{rmclean3d}, which is typically used to deconvolve the rotation measure spread function (RMSF) from the Faraday dispersion function (FDF), thereby reducing sidelobes and mitigating artificial structure. However, since our data do not exhibit significant sidelobe contamination or foreground emission, this step is not necessary in our case.
We measured an RMSF FWHM of $52.33\,\,\mathrm{rad}\,\mathrm{m}^{-2}$, close to the theoretical value of $48.68\,\,\mathrm{rad}\,\mathrm{m}^{-2}$. The theoretical maximum Faraday depth (FD) sensitivity was $1679\,\,\mathrm{rad}\,\mathrm{m}^{-2}$, with the largest scale probed at $96.21\,\,\mathrm{rad}\,\mathrm{m}^{-2}$. The first sidelobe occurs at $53.85\,\,\mathrm{rad}\,\mathrm{m}^{-2}$ with an amplitude of 31.3\,\% of the peak with uniform weights per channel and a uniform noise across the channels. 
The polarisation angle $\chi$ was determined from the Stokes parameters using $\chi = \frac{1}{2} \arctan\left(\frac{U}{Q}\right)$, providing the orientation of the electric field vectors in the plane of the sky. Under the assumption that the observed RM and polarised emission originate from the same region, the vectors were derotated to recover the intrinsic magnetic field orientation via $\chi_0 = \chi - \text{RM} \, \lambda_0^2 + \frac{\pi}{2}$, where $\lambda_0^2$ is the weighted mean of the observed $\lambda^2$ values. For these data, we find $\lambda_0^2 = 0.063\,\mathrm{m}^2$, corresponding to $1.194$\,GHz, which was used as the reference wavelength in the derotation.

\begin{figure}
    \centering
    \includegraphics[width=\linewidth]{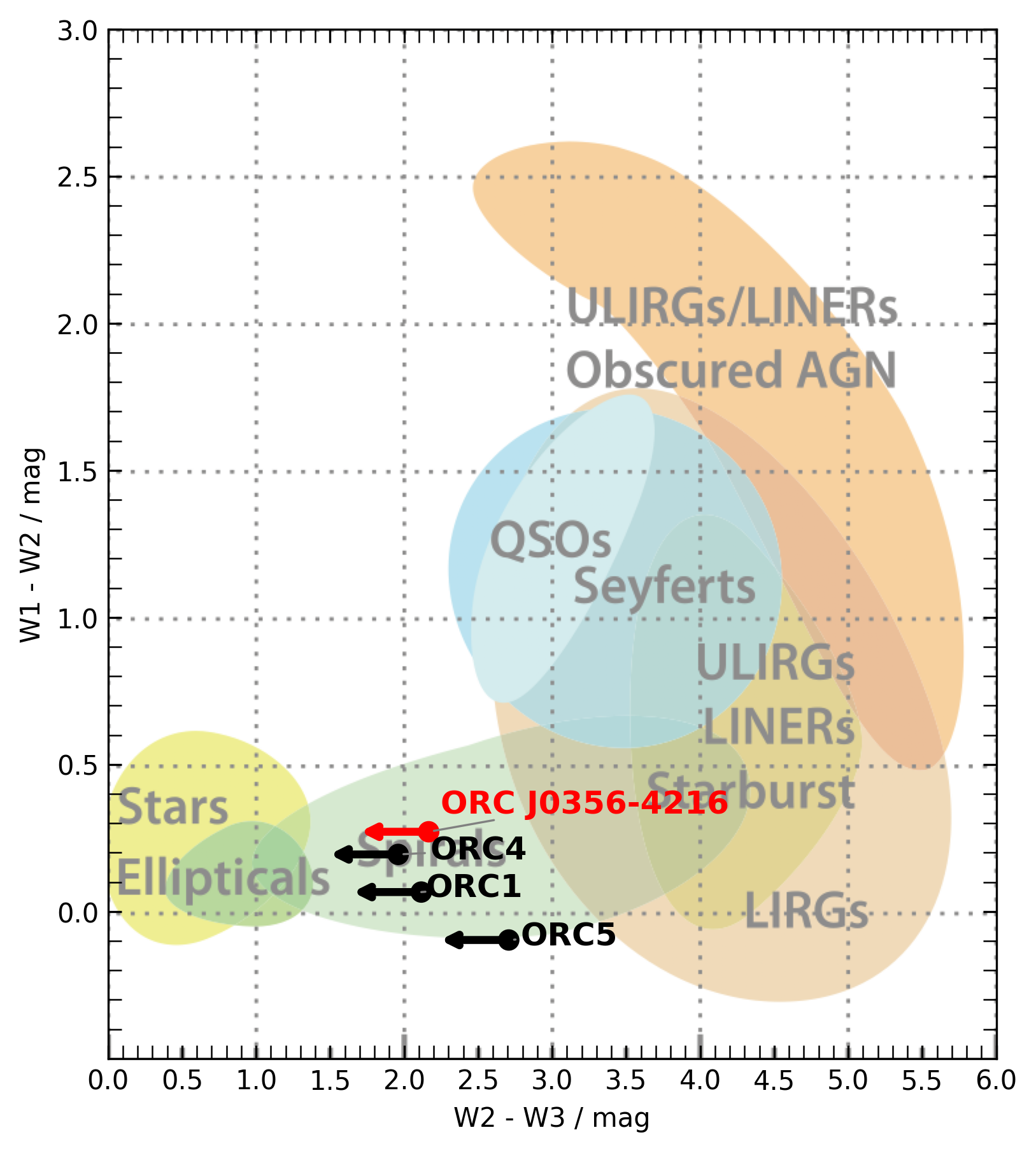}
    \caption{WISE color–color diagram showing in the background the locations of classes of objects \citep[][]{Wright_2010}, while showing the different detected ORCs. The WISE W1, W2 and W3 bands are $3.4\,\upmu$m, $4.6\,\upmu$m and $12\,\upmu$m, respectively. Arrows indicate upper limits.}
    \label{wisecolor}
\end{figure}

\begin{figure*}
    \centering
    \includegraphics[width=\linewidth]{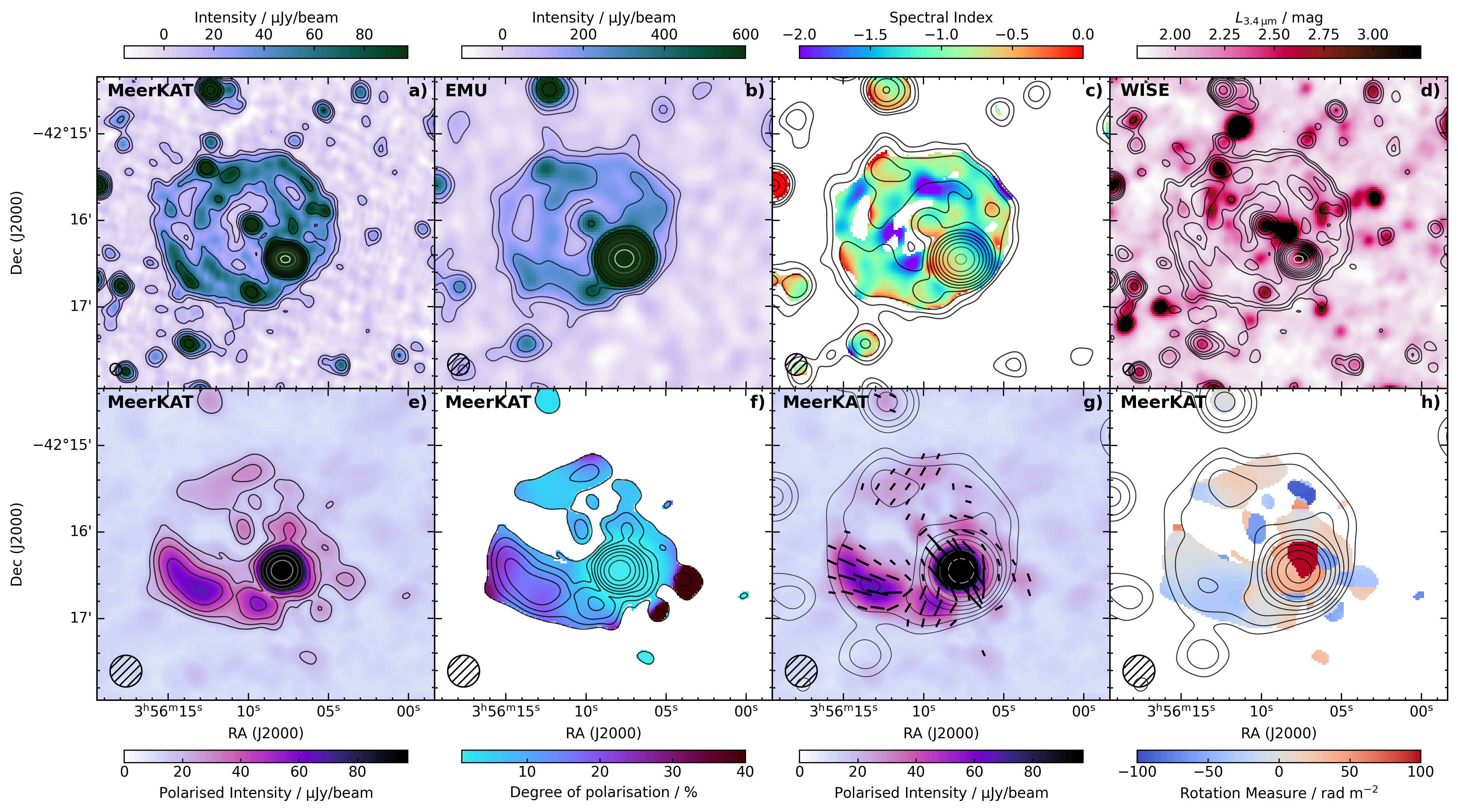}
    \caption{Overview of the relevant radio continuum polarimetry parameters of ORC\,J0356–4216.  
    \textit{Panel a):} Total intensity emission of MeerKAT data at central frequency of 1.28\,GHz and bandwidth of 0.758\,GHz with overlaid contours starting at $3\sigma$ and increasing by factor of two ($\sigma = 3\,\upmu$Jy/beam). The beam of $8.57\arcsec \times 7.65\arcsec$ appears in the lower left corner. 
    \textit{Panel b):} Total intensity emission of EMU data at central frequency of 943\,MHz and bandwidth of 0.288\,MHz with overlaid contours starting at $3\sigma$ and increasing by a factor of two ($\sigma = 20\,\upmu$Jy/beam). A circular beam of $15\arcsec$ appears in the lower left corner. 
    \textit{Panel c):} Two-point total spectral index map between the EMU data of 943\,MHz and MeerKAT data of 1280\,MHz with overlaid contours starting at $3\sigma$ and increasing by a factor of two ($\sigma = 5.5\,\upmu$Jy/beam). The circular beam of $15\arcsec$ appears in the lower left corner. 
    \textit{Panel d):} WISE $3.4\,\upmu$m map with overlaid contours of the MeerKAT 1.28\,GHz emission, starting at $3\sigma$ and increasing by factor of two ($\sigma = 3\,\upmu$Jy/beam). The beam of $8.57\arcsec \times 7.65\arcsec$ appears in the lower left corner.  
    \textit{Panel e):} Polarised intensity emission of MeerKAT data at central frequency of 1.28\,GHz. The contours drawn make use of the signal-to-noise map, starting then at a factor 7 and increase by $\sqrt{2}$. The circular beam of $22\arcsec$ appears in the lower left corner. 
    \textit{Panel f):} Fractional polarisation of MeerKAT data at central frequency of 1.28\,GHz with the same overlaid contours from the polarised intensity map. The contours drawn make use of the signal-to-noise map, starting then at a factor 7 and increase by $\sqrt{2}$. The circular beam of $22\arcsec$ appears in the lower left corner. 
    \textit{Panel g):} Polarised intensity emission of MeerKAT data at central frequency of 1.28\,GHz with overlaid total intensity emission at 1.28\,GHz contours starting at $3\sigma$ and increasing by a factor of two ($\sigma = 30\,\upmu$Jy/beam). The  magnetic field orientations are shown in black and scale with the polarised intensity. The circular beam of $22\arcsec$ appears in the lower left corner.   
    \textit{Panel h):} Distribution of foreground corrected rotation measure of MeerKAT data at central frequency of 1.28\,GHz with overlaid total intensity emission at 1.28\,GHz contours starting at $3\sigma$ and increasing by a factor of two ($\sigma = 30\,\upmu$Jy/beam). The circular beam of $22\arcsec$ appears in the lower left corner.}
    \label{overview}
\end{figure*}
\subsection{EMU}
We detect our ORC-like structure also in the ASKAP EMU main survey observed at a central frequency at 943\,MHz with bandwidth of 288\,MHz \citep[][]{Hopkins_2025}, which is a 10\,hrs integration time observation. We obtained the fully calibrated ASKAP radio continuum image from the CSIRO ASKAP Science Data Archive
(CASDA)\footnote{\url{https://data.csiro.au/domain/casdaObservation}}. Within this field, we reach a local rms value of $20\,\upmu$Jy/beam with a resolution of $15\arcsec$.

\section{Properties of the ORC-like structure}
\label{results}
We propose that the newly discovered object seen in Fig.~\ref{desi}, located at RA $= 03^\text{h}\,56^\text{m}\,09.8^\text{s}$ and Dec $= -42^\text{d}\,16^\text{m}\,03.7^\text{s}$ with a Galactic latitude of $-49.76^\circ$, is a new ORC (ORC\,J0356--4216).   
\subsection{Radio continuum properties}
An analysis of the MeerKAT 1.28\,GHz radio continuum emission reveals a distinct double-ring structure, seen in the right panel of Fig.~\ref{desi} and in panel a) of Fig.~\ref{overview}. This double-ring structure is also visible in the 943\,MHz EMU data, seen in panel b) of Fig.~\ref{overview}. We present a higher resolution image of the ORC\,J0356--4216 in Fig.~\ref{highresol-1} in Appendix~\ref{highresol}, where the double ring structure is even more visible.  Within one of these rings, we identify a bright radio source SUMSS\,J035607--421626, marked in the right panel of Fig.~\ref{desi}. The double ring structure extends over $2^\prime$, corresponding to 667.6\,kpc associated with the host galaxy. ORC\,J0356–4216 fulfills all defining criteria of an ORC, including its exclusive visibility at radio wavelengths, its large diameter spanning several hundred kiloparsecs, and its characteristic ring-like morphology with a potential host galaxy at its center.

\subsection{General Properties}
Based on the position of the radio emission, we assume that the host galaxy is WISEA\,J035609.67--421603.5 of this ORC-like structure, being a massive elliptical galaxy with a photometric redshift of $z_\text{phot} = 0.494\pm0.068$ \citep[][]{Zhou_2021}. The luminosity at 1.4\,GHz of this host galaxy appears to be $\approx3.7\times 10^{23}\,{\rm W\,Hz^{-1}}$ \citep[][]{Sanders_1996} with a starformation rate of about $190\,\mathrm{M_{\odot} \, \mathrm{yr}^{-1}}$ \citep[][]{Bell_2003}, typical of (U)LIRGs.
Fig.~\ref{wisecolor} shows that all host galaxies of the discovered ORCs are not detected in W3 band. Therefore, they should be located in the typical range of elliptical galaxies. In fact, looking at the morphology and optical colors of these host galaxies, it is clear that they are all ellipticals, which is consistent with the WISE upper limit. The signal-to-noise of WISEA\,J035609.67--421603.5 for W3 band is 0.6 in the ALLWISE catalog \citep[][]{Wright_2019}.  Furthermore, we do not perform a K-correction to the WISE colors. Following panel d) of Fig.~\ref{overview} showing the $3.4\,\upmu$m WISE map, we do not see any over-densities within the ring itself, suggesting that the radio emission is not caused by background sources. As shown in Fig.~\ref{desi}, this bright radio emission does not originate from the optical bright source but rather from a more distant galaxy SUMSS\,J035607--421626 with a higher redshift \citep[$z_\text{phot} = 1.320 \pm 0.167$;][]{Zhou_2021}.
This optical bright source revealed to be a star GALEXMSC\,J035612.37--421545.9.

\begin{figure*}
    \centering
    \includegraphics[width=\linewidth]{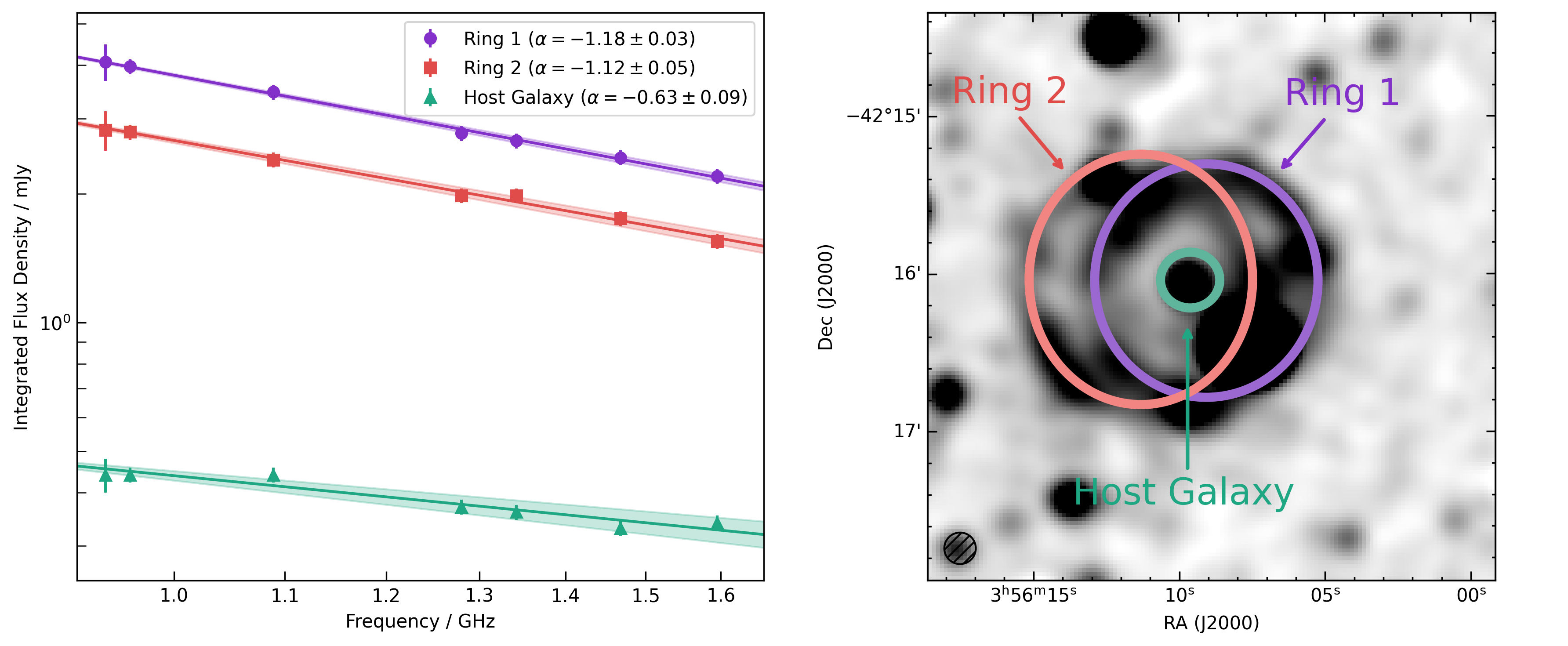}
    \caption{\textit{Left panel:} Fitted spectra of the different components of ORC\,J0356--4216, including ring 1, ring 2, and the host galaxy. The plotted data correspond to the EMU data and the six sub-bands of the MeerKAT observations. We assumed 10\,\% integrated flux density uncertainty for the EMU data and 4\,\% for the MeerKAT data. \textit{Right panel:} MeerKAT 1.28\,GHz background map showing the different components of the galaxy.}
    \label{sed}
\end{figure*}

\subsection{Spectral Analysis}
The total integrated flux density of the rings are 6.67\,mJy for the EMU 943\,MHz and 4.75\,mJy for the MeerKAT 1.28\,GHz, resulting in a spectral index value of $-1.11$. For the spectral index analysis, we selected the MeerKAT 1.28\,GHz and EMU 943\,MHz maps, and clipped both at the $3\,\sigma$ level. We then performed a two-point spectral index analysis between the two maps. To ensure the robustness of the spectral index map, we excluded all values with uncertainties greater than 0.3 of the spectral index error map. The resulting map is shown in panel c) of Fig.~\ref{overview}. The spatial distribution of the spectral index ranges from $-1.7$ to $-0.2$. No clear substructure is observed within the rings.

In Fig.~\ref{sed}, we present the integrated spectrum of each ring and the host galaxy of the ORC\,J0356–4216. To determine the integrated flux density of each ring, we applied masks that exclude the bright radio source SUMSS\,J035607--421626 located in the southwestern region of both rings. These masks were carefully defined as polygons using CARTA \citep[][]{carta} and applied to the MeerKAT and EMU data. The remaining integrated flux density of ring 1 and ring 2 are 2.77\,mJy and 1.98\,mJy, respectively, at a central frequency of 1.28\,GHz. At a central frequency of 943\,MHz for the EMU data, we observe 4.07\,mJy and 2.82\,mJy for ring 1 and ring 2, respectively. 
For this spectral analysis, the MeerKAT data were split into six sub-bands, ranging from 963\,MHz to 1595\,MHz.
The integrated flux density measurements from these sub-bands of MeerKAT and the EMU data are shown in Fig.~\ref{sed}. We assume an uncertainty of 10\,\% for the integrated flux density for the EMU data and 4\,\% for the MeerKAT data. 

The total integrated spectral index of the host galaxy is $-0.63 \pm 0.10$, typical of radio continuum sources. For both rings, we see a spectral index of $-1.18 \pm 0.03$ and $-1.12 \pm 0.05$, respectively, suggesting an aged synchrotron electron population, where radiative losses, primarily due to synchrotron and inverse Compton processes, have steepened the spectrum. The observed value suggests a relatively evolved emission region, with limited recent particle acceleration. The electron energy distribution \citep[$N(E) \propto E^{-p}$ with $p = 1-2\alpha$;][]{Ruszkowski_2023} corresponds to a power-law index of approximately $3.3$, implying significant possible cooling effects. But this may not be entirely true. The injection spectrum depends on the Mach number, and drawing analogy with cluster relics, unlike supernovae in ISM, Mach numbers of about 2--3 in IGM-type densities can also give rise to such type of injection spectra for active CRE acceleration. According to diffusive shock acceleration mechanism, the injection spectral index ($\alpha_{\rm inj}$) is related to Mach number $\mathcal{M}$ as $\alpha_{\rm inj} = -(\mathcal{M}^2 + 3)/(2\mathcal{M}^2 -2)$ \citep{Blandford1987}. This implies, for $\mathcal{M} \approx 2$, $\alpha_{\rm inj} = -1.17$, i.e., $p = 3.34$, what have been observed for the rings. Transonic shocks could also give rise to the observed spectral indices of the rings, which means, a broader frequency coverage is necessary to rule out whether these observations are probing the radiative-loss part of the spectrum, or whether the spectrum continues as a power-law at low frequencies, or steepens towards even higher frequencies.

We now estimate the total magnetic field strength assuming energy equipartition between magnetic fields and cosmic ray particles \citep[][]{Beck_2005}. We note, however, that significant uncertainties are associated with this assumption \citep[e.g.][]{Pfrommer_2004, Seta_2019, Ruszkowski_2023}.  
We use the surface brightness of each ring with its spectral indices, taken from Fig.~\ref{sed}. 
The proton-to-electron energy density ratio was set to $K_0 = 1$ for each ring, in line with values for lobes of radio galaxies \citep[e.g.][]{Adebahr_2019}. Since we lack the necessary correction factor for thermal emission, we assumed that the total radio emission of the rings is equivalent to their synchrotron emission. Additionally, we assume a face-on galaxy, we adopted an inclination angle of $0^\circ$ and a path length of 93\,kpc for the rings, corresponding to their physical thickness, derived from the angular thickness ($13\arcsec$) of the rings relative to the host galaxy.

Under these assumptions, we calculated the total equipartition magnetic field strength for the two rings. For ring 1, the total equipartition magnetic field strength is $1.82\,\upmu$G, while for ring 2, the total equipartition magnetic field strength is estimated to $1.65\,\upmu$G for a rest-frequency of 1.92\,GHz. These values are higher found in radio lobes in \citep[][]{Adebahr_2019} and also much smaller compared to massive edge-on galaxies \citep[e.g.][]{Stein_2023} or dwarf galaxies \citep[e.g.][]{Basu_2017, Taziaux_2025}. 
However, these equipartition magnetic field strengths are significantly below $3.25\,(1+z)^2 = 7.25\,\upmu$G, therefore the rings should be undergoing inverse-Compton losses. 

\subsection{Polarisation Properties}
We do not expect significant off-axis leakage in L-band at $6.68\arcmin$ from the phase centre \citep[e.g.][]{Sekhar_2022}. In the bottom row of Fig.~\ref{overview}, we present polarisation-related parameters only for MeerKAT, i.e., polarised intensity, fractional polarisation, magnetic field orientations, and the rotation measure of ORC\,J0356--4216. 
To avoid Ricean bias that originates from positive definite background in polarised intensity map, we clipped above $7\,\sigma_{\rm PI}$ to ensure only taking true emission into account. For this, we estimate the uncertainties from a slice from the Stokes$-Q$ and $-U$ map, which is located at a Faraday depth farer away then from the typical depth of the ORC\,J0356--4216 ($\sigma_{Q}=2.53\,\upmu$Jy/beam; $\sigma_{U}=2.74\,\upmu$Jy/beam). The polarised intensity uncertainties have been calculated through error propagation, resulting in $\sigma_{\rm PI} = \sqrt{\frac{Q^2}{Q^2+U^2}\sigma_Q+\frac{U^2}{Q^2+U^2}\sigma_U}$. The contours drawn at the polarisation intensity map in panel e) in Fig.~\ref{overview}, we use the signal-to-noise map, starting then at a factor $7$ and increase by $\sqrt{2}$. 
As the effective frequency of the polarisation emission map, generated by RM synthesis, is not at the central frequency but at the weighted mean $\lambda^2_0$, we normalise the Stokes$-I$ to 1.194\,GHz to generate the fractional polarisation map in panel f) in Fig.~\ref{overview}.

We detect high polarised intensity in the radio source located within the ring, although the fractional polarisation is below $10\,\%$. The polarised intensity of the ring alone, excluding the contribution from the radio source SUMSS\,J035607–421626, is measured to be 0.47\,mJy. In comparison, the polarised intensity of the bright radio source in the ring by itself is 0.23\,mJy. In contrast to the bright radio source, the left part of the ring shows a much higher degree of polarisation, ranging from $20\,\%$ to $30\,\%$. The northern part of the structure has a polarisation degree between $10\,\%$ and $20\,\%$. This suggests that the ring itself is the primary source of the observed polarised emission in the system, with its higher degree of polarisation and greater integrated polarised intensity outweighing the contribution from the embedded radio source.
The magnetic field orientation, corrected for RM, appear to follow the general shape of the ring, but their orientation is not perfectly aligned with the ring, having a small offset with respect to the total intensity.
It can be described as they point tangential to the rings. The rotation measure values range from $-93\,\text{rad\,m}^{-2}$ to $110\,\text{rad\,m}^{-2}$. These values have been corrected for the Milky Way foreground \citep[][]{Hutschenreuter_2022}. The rotation measure value of the Galactic foreground is $3.54 \pm 0.92\,\text{rad\,m}^{-2}$, which is relatively low. Positive values represent a magnetic field that is pointing towards us, while negative ones describe the opposite. The left side of the ring predominantly shows negative values, while the northern region mostly displays positive values.
This could suggest that we have here a toroidal field. 

\section{Discussion}
\label{discussion}
\subsection{Environment of the ORC\,J0356--4216}
\begin{figure*}
    \centering
\includegraphics[width=\linewidth]{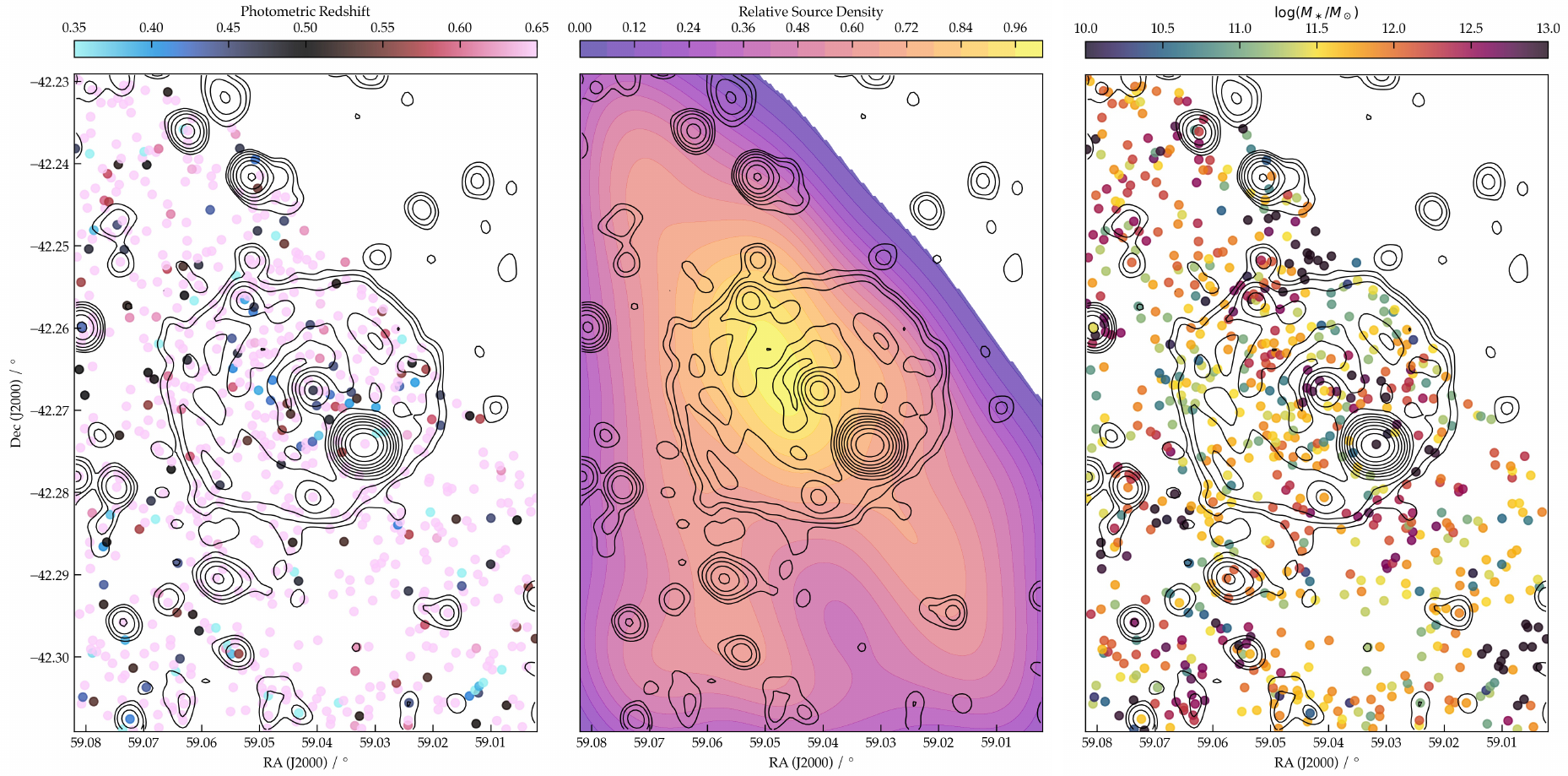}
    \caption{\textit{Left panel:} Photometric redshift distribution of all the sources around ORC\,J0356--4216 taken from Legacy Survey DR9 photo-z \citep[][]{Zhou_2021}. All galaxies, below a photometric redshift of 0.35 are shown in light-blue and all galaxies above redshift 0.65 are displayed in pink. \textit{Middle panel:} Kernel density estimate map of galaxy positions in the same field, showing the relative projected source density, based on the source information from the Legacy Survey DR9. \textit{Right panel:} Spatial distribution of stellar masses for galaxies in the field, derived from WISE $W1$ photometry and photometric redshifts, using the stellar mass calibration from \citet{Jarrett_2023}. }
    \label{evironments}
\end{figure*}
To better understand the nature of the ORC\,J0356--4216, we examine the distribution of its radio continuum emission by analysing the area around the host galaxy and its rings to see if there is any noticeable overdensity. In panel d) of Fig.~\ref{overview}, we look at the WISE $3.4\,\upmu\text{m}$ magnitude, which is used as an indicator of stellar mass. Their redshift distribution is shown in left panel of Fig.~\ref{evironments}, where we see that the host galaxy is at a redshift of $0.494 \pm 0.068$, with a few sources at similar redshifts, showing a connection to the host, see in the $3.4\,\upmu\text{m}$ magnitude in panel d) of Fig.~\ref{overview}), which could point to the galaxy merger scenario. Several galaxies, with redshifts between $0.4 < z_\text{phot} < 0.6$, are located within $0.5\arcmin$ of the host galaxy. Within the region of radio emission in the rings, there are several galaxies in a similar redshift range. This suggests that the source is located in an overdense environment, similar to  ORC\,1 \citep[][]{Norris_2021}.

To investigate the environment of ORC\,J0356--4216 in more detail, we show a kernel density estimate in the middle panel of Fig.~\ref{evironments} to compare the density of galaxies inside and outside the ORC. The plot indicates that there are more galaxies within the ORC than in the surrounding area, suggesting a higher density inside the structure.

Since we observe an overdensity within the ring structure (middle panel of Fig.\ref{evironments}), we examine whether the radio emission originates from the galaxy WISEA J035609.67--421603.5. For this purpose, we present high-resolution images in Appendix~\ref{highresol}, created using a more uniform Briggs robust weighting to improve the resolution. In Fig.~\ref{highresol-2}, we show a high-resolution image of ORC\,J0356--4216 overlaid on an optical image. This comparison indicates that the elliptical galaxy WISEA J035609.67--421603.5 is the likely host galaxy of ORC\,J0356--4216.

In the right panel of Fig.~\ref{evironments}, we show the stellar mass distribution in the environment of ORC\,J0356--4216. The masses are estimated using the redshift and 3.4\,$\upmu$m magnitude from WISE \citep[Eq.~2,][]{Jarrett_2023}. For galaxies with $>10^9\,$M$_\odot$, the derived stellar mass estimates have an accuracy of about $25$–$30\,\%$. No K-correction was applied. For ORC\,J0356--4216, we find a stellar mass of log(M$_*$/M$_\odot$) = 11.66. For comparison, we calculate stellar masses of 12.26 for ORC\,1, 11.59 for ORC\,4, and 11.03 for ORC\,5, all within a similar range than ORC\,J0356--4216. These host galaxies are all massive ellipticals, consistent with the idea that such large ring-like structures are associated with massive galaxies.

\subsection{Physical nature of ORC\,J0356--4216}
There are several hypotheses regarding the origin and nature of ORCs, but none have been definitively confirmed. Here, we only discuss the physical nature of ORC\,J0356–4216.
Based on the structure of the radio continuum emission, the ORC\,J0356–4216 appears to have the shape of a shell or a shock front. A supernova remnant shock in the Milky Way, as suggested by \citet[][]{Filipovic_2022}, is unlikely in this case as the host galaxy is located at a galactic latitude of $-49.76^\circ$, far from the dense Galactic disk, where SNRs are typically found.

Another possibility is that the ORC originates from a large-scale shock within the host galaxy, potentially triggered by a powerful burst of star formation. Such an event could generate outward-propagating shock waves that sweep up surrounding material into ring-like shells, as proposed by \citet[][]{Yamasaki_2024}.

The double-ring structure seen in ORC\,J0356–4216 closely resembles the appearance of radio lobes produced by an AGN, where jets are ejected in opposite directions. Additionally, the orientation of the magnetic fields, which generally follow the ring structure with a slightly offset angle, is consistent with what is expected from AGN jet models. \citet[][]{Lin_2024} proposed that ORCs may originate from end-on cosmic ray proton-dominated AGN bubbles. The AGN jets are powerful enough to clear out high-density gas in lower-mass clusters or groups, which could host ORCs. High hadronic collision rates at the bubble surface could produce secondary particles emitting synchrotron radiation, resulting in limb-brightened radio morphology. 

To investigate whether the ORC could indeed be related to AGN activity, we plotted a WISE color-color diagram in Fig.~\ref{wisecolor}. Most AGN and radio galaxies are  ellipticals with W2-W3 $\approx 1.0$, consistent with the colours of this host galaxy.
However, more detailed observations, such as optical spectroscopy and the use of a BPT diagram, would be necessary to confirm the absence or presence of AGN activity.
When we compare our ORC with other confirmed ORCs in Fig.~\ref{wisecolor}, we find that most of them occupy a similar region in the WISE color-color space, characterised by low W2–W3 and W1–W2 colors. An exception is ORC\,2/3, which has been proposed to be associated with AGN lobes \citep[][]{Koribalski_2021}.

Alternatively, we may be observing a remnant of an AGN that is currently fading. \citet[][]{Shabala_2024} explores the remnant phase of evolution after the jets have been switched off, finding that the synchrotron emission fades rapidly during this phase. This may result from bipolar outflows, similar to what \citet[][]{Coil_2024} reported for ORC4, and further supported by the magnetic field observations in this ORC.
Additionally, the simulations investigate the re-acceleration of cosmic ray electrons through the passage of a plane parallel shock, leading to a `phoenix' phase of radio emission.

Another hypothesis is that ORCs might be related to bent radio galaxies, where the galaxies are bent by interactions with the surrounding medium. The bright radio source SUMSS J035607--421626 may suggest that the system is a wide-angle tail radio galaxy, particularly given the arc-like morphology of the high-intensity emission observed in Fig.~\ref{overview} a) and b). 
However, this scenario is unlikely for ORC\,J0356--4216, as we do not observe any substructures in the spatial distribution of the spectral index. In the bent galaxy scenario, we would expect the spectral index to steepen progressively toward the ends of the galaxy, which is not observed here.

We propose two possible scenarios for the formation of this ORC: (1) it may have originated from galactic shocks in the massive host galaxy, potentially triggered by a powerful burst with a bipolar jet, as supported by the morphology of the radio emission and polarisation; and (2) it is possible that this ORC is connected to AGN jets forming extended radio lobes, or may represent the remnant of past AGN activity, particularly given its double-ring structure.\\

\section{Conclusions}
\label{conclusion}
We present the first radio continuum polarimetry study of a new ORC-like system, which we dub ORC\,J0356--4216. The deep MeerKAT observation reveals an double ring structure which could be associated with an ORC and which leads us to the following conclusions
\begin{enumerate}
    \item We detect radio continuum emission of a double ring structure, which cannot be detected in optical, infrared or other wavelengths. We detect radio continuum emission in the ASKAP survey EMU at 943\,MHz and MeerKAT 1.28\,GHz observations, which can be associated with being an ORC. This ORC is located at a redshift of $0.494\pm0.068$ and has a diameter of $2\arcmin$, corresponding to 667\,kpc according to the host galaxy.  
    \item The integrated flux density for the EMU data at 943\,MHz of the rings is 4.07\,mJy and 2.82\,mJy, repsectively.
    The integrated flux density of each ring at a central frequency of 1.28\,GHz is 2.77\,mJy and 1.98\,mJy possessing a steep spectrum with a slope of $-1.18 \pm 0.03$ and $-1.12 \pm 0.05$, respectively, suggesting synchrotron and inverse Compton losses being the dominant factor. 
    Assuming purely synchrotron emission at these frequencies, the total equipartition magnetic field reaches values of $1.82\,\upmu$G and $1.65\,\upmu$G, respectively.
    \item The polarised intensity of the rings of the ORC are 0.47\,mJy, corresponding to a degree of polarisation of 20 to 30\,\% through the rings. The magnetic field orientations follow the rings pointing tangential to the rings. 
    \item The origin of ORCs remains uncertain, with two main scenarios under discussion for ORC\,J0356--4216: a remnant from a past AGN phase or the result of a large-scale shock wave, such as those generated by galaxy group interactions or mergers. In the case of ORC\,J0356--4216, the observed double-lobed morphology and polarisation characteristics are more readily explained by relic emission from previous AGN activity or jet-driven outflows. The WISE colours are consistent with the host being an AGN in an elliptical galaxy. Given the available data, an AGN-related origin appears to be the more consistent interpretation.
\end{enumerate}

\section{Outlook}
Estimating the volumetric number density of ORCs is challenging, as they are fainter than massive grand-design galaxies and even many dwarf galaxies, requiring high sensitivity for detection. In addition to sensitivity, good $uv$-coverage and high angular resolution are also essential to recover their diffuse emission. Based on the currently published detections, there are roughly 10 securely identified ORCs, with a few additional candidates. Most of these have been found in wide-field radio surveys such as the ASKAP EMU Pilot Survey ($\sim270$\,deg$^2$ at full sensitivity), supplementary ASKAP observations, GMRT ($\sim$10 to 100\,deg$^2$). These surveys collectively cover an effective area of approximately 1000\,deg$^2$, sensitive to ORCs out to redshifts of about $z \lesssim 0.6$. This corresponds to a comoving survey volume of roughly 5 Gpc$^3$. From these numbers, we estimate a present-day volumetric number density of $n \approx (2 \pm 1) \, \mathrm{Gpc}^{-3}$, where the uncertainty reflects both Poisson statistics and the approximate nature of the survey volume. This estimate is subject to significant biases and incompleteness. ORCs are extended, low-surface-brightness sources, so their detection strongly depends on survey depth, angular resolution, and frequency coverage. In addition, most have been found via visual inspection, which favours the discovery of symmetric, "classic" ring morphologies and may overlook irregular or partial structures. Redshift incompleteness also affects volume estimates, as not all ORCs have secure host identifications. Consequently, the quoted density should be regarded as a lower limit, with the true space density potentially being higher.

\bibliographystyle{aa} 
\bibliography{literatur} 

\begin{acknowledgements}
ST, BA, CJR and DJB acknowledge the support from the DFG via the Collaborative Research Center SFB1491 \textit{Cosmic Interacting Matters - From Source to Signal}.
This scientific work uses data obtained from Inyarrimanha Ilgari Bundara, the CSIRO Murchison Radio-astronomy Observatory. We acknowledge the Wajarri Yamaji People as the Traditional Owners and native title holders of the Observatory site. CSIRO’s ASKAP radio telescope is part of the Australia Telescope National Facility (https://ror.org/05qajvd42). Operation of ASKAP is funded by the Australian Government with support from the National Collaborative Research Infrastructure Strategy. ASKAP uses the resources of the Pawsey Supercomputing Research Centre. Establishment of ASKAP, Inyarrimanha Ilgari Bundara, the CSIRO Murchison Radio-astronomy Observatory and the Pawsey Supercomputing Research Centre are initiatives of the Australian Government, with support from the Government of Western Australia and the Science and Industry Endowment Fund.
This paper includes archived data obtained through the CSIRO ASKAP Science Data Archive, CASDA (http://data.csiro.au).
We acknowledge data storage and computational facilities by the University of Bielefeld which are hosted by the Forschungszentrum Jülich and that were funded by German Federal Ministry of Education and Research (BMBF) projects D-LOFAR IV (05A17PBA) and D-MeerKAT-II (05A20PBA), as well as technical and operational support by BMBF projects D-LOFAR 2.0 (05A20PB1) and D-LOFAR-ERIC (05A23PB1), and German Research Foundation (DFG) project PUNCH4NFDI (460248186). 
The Photometric Redshifts for the Legacy Surveys (PRLS) catalog used in this paper was produced thanks to funding from the U.S. Department of Energy Office of Science, Office of High Energy Physics via grant DE-SC0007914.
The Legacy Surveys consist of three individual and complementary projects: the Dark Energy Camera Legacy Survey (DECaLS; Proposal ID \#2014B-0404; PIs: David Schlegel and Arjun Dey), the Beijing-Arizona Sky Survey (BASS; NOAO Prop. ID \#2015A-0801; PIs: Zhou Xu and Xiaohui Fan), and the Mayall z-band Legacy Survey (MzLS; Prop. ID \#2016A-0453; PI: Arjun Dey). DECaLS, BASS and MzLS together include data obtained, respectively, at the Blanco telescope, Cerro Tololo Inter-American Observatory, NSF’s NOIRLab; the Bok telescope, Steward Observatory, University of Arizona; and the Mayall telescope, Kitt Peak National Observatory, NOIRLab. Pipeline processing and analyses of the data were supported by NOIRLab and the Lawrence Berkeley National Laboratory (LBNL). The Legacy Surveys project is honored to be permitted to conduct astronomical research on Iolkam Du’ag (Kitt Peak), a mountain with particular significance to the Tohono O’odham Nation.
NOIRLab is operated by the Association of Universities for Research in Astronomy (AURA) under a cooperative agreement with the National Science Foundation. LBNL is managed by the Regents of the University of California under contract to the U.S. Department of Energy.
This project used data obtained with the Dark Energy Camera (DECam), which was constructed by the Dark Energy Survey (DES) collaboration. Funding for the DES Projects has been provided by the U.S. Department of Energy, the U.S. National Science Foundation, the Ministry of Science and Education of Spain, the Science and Technology Facilities Council of the United Kingdom, the Higher Education Funding Council for England, the National Center for Supercomputing Applications at the University of Illinois at Urbana-Champaign, the Kavli Institute of Cosmological Physics at the University of Chicago, Center for Cosmology and Astro-Particle Physics at the Ohio State University, the Mitchell Institute for Fundamental Physics and Astronomy at Texas A\&M University, Financiadora de Estudos e Projetos, Fundacao Carlos Chagas Filho de Amparo, Financiadora de Estudos e Projetos, Fundacao Carlos Chagas Filho de Amparo a Pesquisa do Estado do Rio de Janeiro, Conselho Nacional de Desenvolvimento Cientifico e Tecnologico and the Ministerio da Ciencia, Tecnologia e Inovacao, the Deutsche Forschungsgemeinschaft and the Collaborating Institutions in the Dark Energy Survey. The Collaborating Institutions are Argonne National Laboratory, the University of California at Santa Cruz, the University of Cambridge, Centro de Investigaciones Energeticas, Medioambientales y Tecnologicas-Madrid, the University of Chicago, University College London, the DES-Brazil Consortium, the University of Edinburgh, the Eidgenossische Technische Hochschule (ETH) Zurich, Fermi National Accelerator Laboratory, the University of Illinois at Urbana-Champaign, the Institut de Ciencies de l’Espai (IEEC/CSIC), the Institut de Fisica d’Altes Energies, Lawrence Berkeley National Laboratory, the Ludwig Maximilians Universitat Munchen and the associated Excellence Cluster Universe, the University of Michigan, NSF’s NOIRLab, the University of Nottingham, the Ohio State University, the University of Pennsylvania, the University of Portsmouth, SLAC National Accelerator Laboratory, Stanford University, the University of Sussex, and Texas A\&M University.
BASS is a key project of the Telescope Access Program (TAP), which has been funded by the National Astronomical Observatories of China, the Chinese Academy of Sciences (the Strategic Priority Research Program “The Emergence of Cosmological Structures” Grant \# XDB09000000), and the Special Fund for Astronomy from the Ministry of Finance. The BASS is also supported by the External Cooperation Program of Chinese Academy of Sciences (Grant \# 114A11KYSB20160057), and Chinese National Natural Science Foundation (Grant \# 12120101003, \# 11433005).
The Legacy Survey team makes use of data products from the Near-Earth Object Wide-field Infrared Survey Explorer (NEOWISE), which is a project of the Jet Propulsion Laboratory/California Institute of Technology. NEOWISE is funded by the National Aeronautics and Space Administration.
The Legacy Surveys imaging of the DESI footprint is supported by the Director, Office of Science, Office of High Energy Physics of the U.S. Department of Energy under Contract No. DE-AC02-05CH1123, by the National Energy Research Scientific Computing Center, a DOE Office of Science User Facility under the same contract; and by the U.S. National Science Foundation, Division of Astronomical Sciences under Contract No. AST-0950945 to NOAO.
\end{acknowledgements}

\appendix
\onecolumn
\section{High-resolution images}
\label{highresol}
\begin{figure*}[h!]
    \centering
    \includegraphics[width=0.8\linewidth]{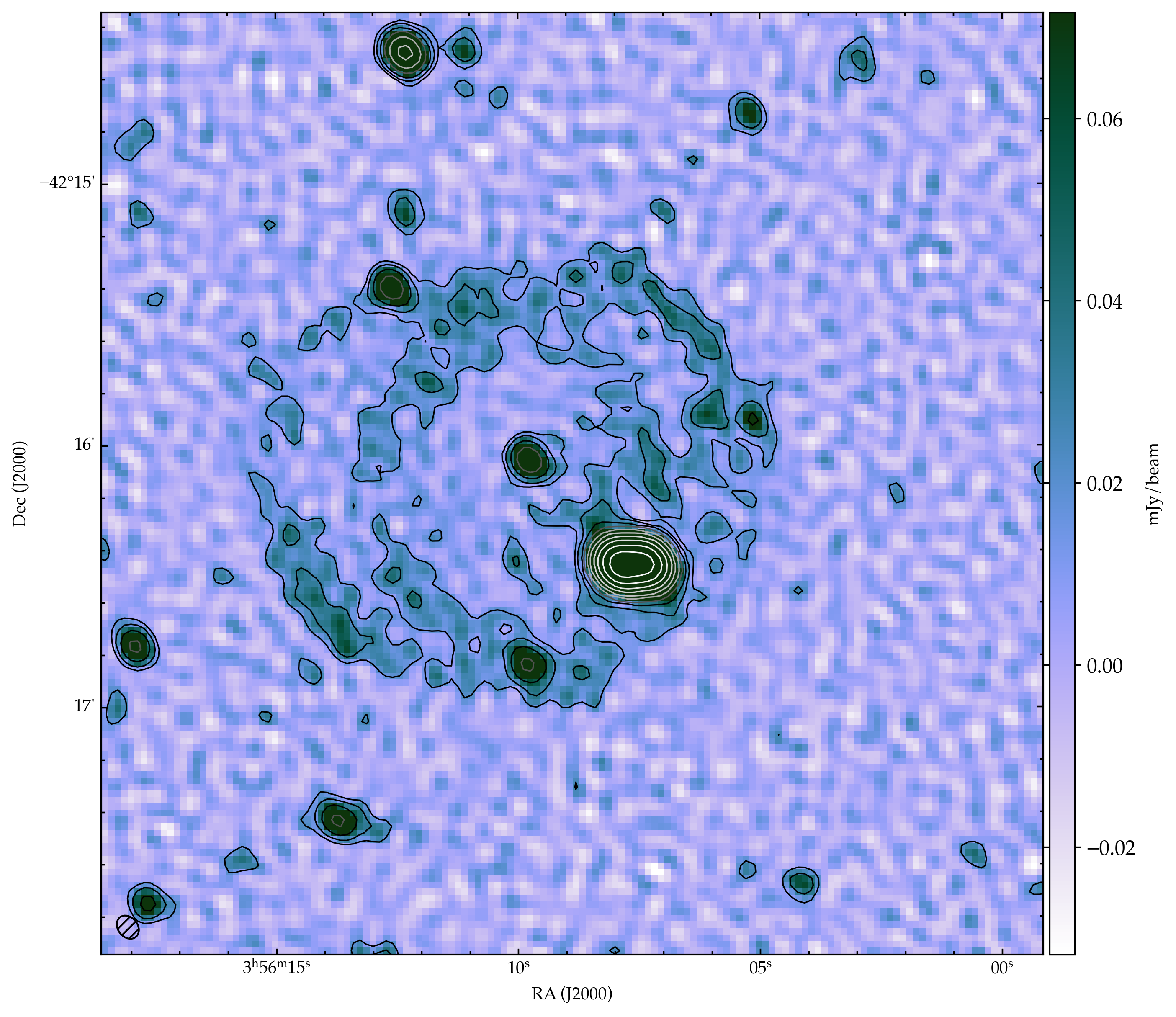}
    \caption{MeerKAT L-band radio emission imaged at \texttt{robust}$ = -1$ with overlaid contours starting at $3\,\sigma$ and increasing by a factor of 2 at a central frequency of 1.28\,GHz with a noise level $\sigma=5\,\upmu$Jy/beam.
    The beam of $5.8\arcsec \times 4.5\arcsec$ is shown in the bottom left corner.}
    \label{highresol-1}
\end{figure*}

\begin{figure*}[h!]
    \centering
     \includegraphics[width=\linewidth]{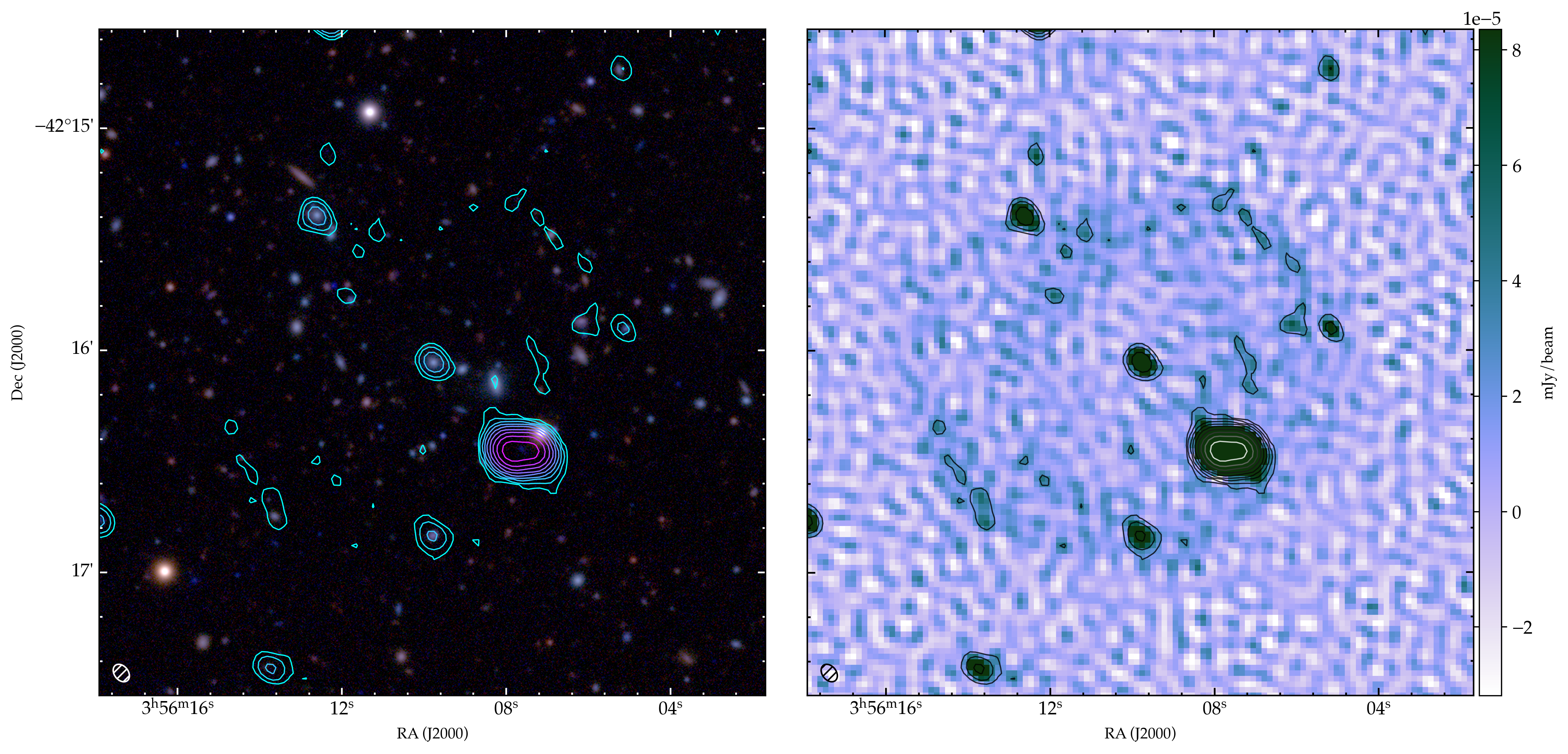}
    \caption{\textit{left:} Color-composite image from the DESI Legacy Imaging Surveys of ORC\,J0356–4216 with overlaid MeerKAT L-band radio emission contours starting at $3\,\sigma$ and increasing by a factor of 2 at a central frequency of 1.28\,GHz with a noise level $\sigma=7.5\,\upmu$Jy/beam.
    \textit{Right:} MeerKAT L-band radio emission imaged at \texttt{robust}$ = -2$ with overlaid contours starting at $3\,\sigma$ and increasing by a factor of 2 at a central frequency of 1.28\,GHz with a noise level $\sigma=7.5\,\upmu$Jy/beam.
    The beam of $5.2\arcsec \times 3.7\arcsec$ is shown in the bottom left corner.}
    \label{highresol-2}
\end{figure*}

High-resolution images of ORC\,J0356–4216 are presented in this section. The MeerKAT data have been processed using the same calibration, self-calibration, and imaging procedures described in Sect.~\ref{datameerkat}, with the exception that the final images were produced using Briggs weighting with a robust parameters of $\texttt{robust} = -1$ presented in Fig.~\ref{highresol-1} and $\texttt{robust} = -2$ in Fig.~\ref{highresol-2}. This results in a sensitivity of $5\,\upmu$Jy/beam and $7\,\upmu$Jy/beam, respectively, and an angular resolution of $5.8\arcsec \times 4.5\arcsec$ and $5.2\arcsec \times 3.7\arcsec$, respectively. The improved resolution allows for a clearer separation of the different components within ORC\,J0356–4216, particularly the ring-like structure and the candidate host galaxy.

A key aspect of this analysis is that the estimates of both luminosity and physical size rely on the assumption that the central peak of the radio emission corresponds to the optical host galaxy. Inspection of the DESI image in Fig.~\ref{desi} reveals the presence of other, redder galaxies in the field; however, given the alignment and morphology of the higher resolution image in Fig.~\ref{highresol-2}, these can likely be excluded as potential hosts.



\label{lastpage}
\end{document}